\documentclass[preprint,authoryear,5p]{elsarticle}

\usepackage{graphicx,amssymb}
\usepackage{epsf,psfig}
\usepackage{natbib}

\journal{Mechanics Research Communications}

\begin{document}

\begin{frontmatter}
\title{Orbit classification of low and high angular momentum stars}

\author{Euaggelos E. Zotos\corref{cor1}}
\ead{evzotos@physics.auth.gr}

\cortext[cor1]{Corresponding author}

\address{Department of Physics, School of Science, \\
Aristotle University of Thessaloniki, \\
GR-541 24, Thessaloniki, Greece \\}

\begin{abstract}
We determine the character of orbits of stars moving in the meridional plane $(R,z)$ of an axially symmetric time-independent disk galaxy model with a spherical central nucleus. In particular, we try to reveal the influence of the value of the angular momentum on the different families of orbits of stars, by monitoring how the percentage of chaotic orbits, as well as the percentages of orbits of the main regular resonant families evolve when angular momentum varies. The smaller alignment index (SALI) was computed by numerically integrating the equations of motion as well as the variational equations to extensive samples of orbits in order to distinguish safely between ordered and chaotic motion. In addition, a method based on the concept of spectral dynamics that utilizes the Fourier transform of the time series of each coordinate is used to identify the various families of regular orbits and also to recognize the secondary resonances that bifurcate from them. Our investigation takes place both in the physical $(R,z)$ and the phase $(R,\dot{R})$ space for a better understanding of the orbital properties of the system. Our numerical computations reveal that low angular momentum stars are most likely to move in chaotic orbits, while on the other hand, the vast majority of high angular momentum stars perform regular orbits.
\end{abstract}

\begin{keyword}
Galaxies: kinematics and dynamics -- Galaxies: structure, chaos
\end{keyword}

\end{frontmatter}

\section{Introduction}
\label{intro}

Knowing the regular or chaotic nature of orbits in galaxies is an issue of paramount importance. This is true because this knowledge allows us to understand and interpret the formation and also predict the evolution of galaxies. In addition, families of regular orbits are often used as the basic tool in constructing a dynamical model for describing the main properties of galaxies. Over the last several decades, a huge amount of research work has been devoted to understanding the orbital structure in different types of galaxy models \citep[e.g.,][]{P84,CG89,SW93,P96,OP98,PMM04}. However, the vast majority of the existing literature deals only either with the distinction between regular and chaotic motion \citep[e.g.,][]{MA11,BMA12,MBS13} or the detection of periodic orbits and the analysis of their stability \citep[e.g.,][]{SPA02a,SPA02b,KP05}. We would like to note that all the above-mentioned references on the dynamics of galaxies are exemplary rather than exhaustive. In the present paper, on the other hand, we proceed one step further contributing to this active field by classifying ordered orbits into different regular families following the steps of \citet{CA98}.

An interesting fact is that only a small fraction of the existing literature deals with motion of stars on the meridional plane of an axially symmetric potential. Investigation of this particular type of motion can be traced back to the studies of \citet{C60} and \citet{O65,O67}. The nature of resonant meridional plane orbits has been explored in \citet{MM75}, while \citet{M79} considered, as was common in those days, that almost any orbit in an axially symmetric potential should obey a third isolating integral of motion besides the angular momentum and the total orbital energy. The motion of stars in the meridional plane has been characterized in \citet{C79}, as one of opening issues in galactic dynamics, where integrability and stochasticity play a vital role. \citet{G87} following the work of \citet{M79}, continued to ignore the chaoticity, while on the other hand, chaos was eventually found in \citet{CV86} but only when a perturbation was added to the galactic model. In the same vein, \citet{GS91} and \citet{CZD00} investigated the dynamics of orbits in the meridional plane, but once more, they focused only on ordered motion suggesting that most stars move on regular orbits. Chaotic motion in a two-dimensional logarithmic potential describing the properties of an elliptical galaxy with a dense bulge has been examined in \citet{KC01}, although it does not contain the necessary centrifugal term. The orbital content of triaxial potentials as well as of axisymmetric potentials has been thoroughly analyzed in the pioneer work of \citet{LS92}.

At this point, we should mention that in our previous work we have also investigated the character of orbits moving in the meridional plane of axially symmetric potentials. A new galactic potential was introduced in \citet{Z11} in order to describe the motion of stars moving in the meridional plane of disk or elliptical galaxies. Furthermore, the role of the central massive nucleus on the character of orbits in the $(R,z)$ plane has been investigated in \citet{Z12}. In \citet{CZ13,ZCar13} new types of axisymmetric potentials have been developed in an attempt to model the properties and elucidate the influence of dark matter in the main galactic body. The nature of orbits of stars moving in the meridional plane of an axially symmetric galactic model with a disk, a spherical nucleus, and a flat biaxial dark matter halo component has been explored in \citet{Z14}. In the same line, in \citet{ZC13} (hereafter Paper I) we used an analytic axisymmetric potential which embraces the general features of a disk galaxy with a bulge, in order to reveal how influential are the main parameters of the system on the level of chaos and on the distribution of regular families of orbits. There is no doubt, that the issue of motion of stars in the meridional plane of axially symmetric galaxies, is still an open and active problem. In the current work, we will continue the investigation started in Paper I but in this case we shall focus to a specific dynamical quantity which is the angular momentum. In particular, our aim is to determine the character of low and high angular momentum stars.

The layout of the paper is as follows: Section \ref{galmod}, contains a detailed presentation of the structure and the properties of our galactic gravitational model. In Section \ref{orbclas}, we investigate how the angular momentum influences the nature as well as the evolution of the percentages of the different families of orbits. The paper ends with Section \ref{disc}, where the main conclusions of our numerical analysis and the discussion are presented.

\section{Presentation of the galactic model}
\label{galmod}

The main objective of this investigation is to reveal the regular or chaotic nature of orbits of low and high angular momentum stars moving in the meridional plane of an axially symmetric disk galaxy with a spherical central nucleus. We use the usual cylindrical coordinates $(R, \phi, z)$, where $z$ is the axis of symmetry.

The total gravitational potential $\Phi(R,z)$ in our model consists of two components: the central spherical component $\Phi_{\rm n}$ and the disk potential $\Phi_{\rm d}$. For the description of the spherically symmetric nucleus, we use a Plummer potential
\begin{equation}
\Phi_{\rm n}(R,z) = \frac{- G M_{\rm n}}{\sqrt{R^2 + z^2 + c_{\rm n}^2}}.
\label{Vn}
\end{equation}
Here $G$ is the gravitational constant, while $M_{\rm n}$ and $c_{\rm n}$ are the mass and the scale length of the nucleus, respectively. This potential has been used successfully in the past to model and, therefore, interpret the effects of the central mass component in a galaxy \citep[see, e.g.,][]{Z12,ZC13,Z14}. At this point, we must emphasize that we do not include any relativistic effects, because the nucleus represents a bulge rather than a black hole or any other compact object. The galactic disk on the other hand, is represented by the well-known Miyamoto-Nagai potential \citep{MN75}
\begin{equation}
\Phi_{\rm d}(R,z) = \frac{- G M_{\rm d}}{\sqrt{R^2 + \left(\alpha + \sqrt{h^2 + z^2}\right)^2}},
\end{equation}
where, $M_{\rm d}$ is the mass of the disk, $\alpha$ is the scale length of the disk, and $h$ corresponds to the disk's scale height.

We use a system of galactic units where the unit of length is 1 kpc, the unit of velocity is 10 km s$^{-1}$, and $G = 1$. Thus, the unit of mass is $2.325 \times 10^7 {\rm M}_\odot$, that of time is $0.9778 \times 10^8$ yr, the unit of angular momentum (per unit mass) is 10 km kpc$^{-1}$ s$^{-1}$, and the unit of energy (per unit mass) is 100 km$^2$s$^{-2}$. In these units, the values of the involved parameters are: $M_{\rm n} = 400$ (corresponding to $9.3\times 10^{9}$ M$_\odot$), $c_{\rm n} = 0.25$, $M_{\rm d} = 7000$ (corresponding to $1.63\times 10^{11}$ M$_\odot)$, $\alpha = 3$ and $h = 0.175$. The particular values of the parameters were chosen with a Milky Way-type galaxy in mind \citep[e.g.,][]{AS91}. The above-mentioned set of values of the parameters which are kept constant throughout the numerical calculations secures positive mass density everywhere and free of singularities.

\begin{figure}[!tH]
\centering
\includegraphics[width=\hsize]{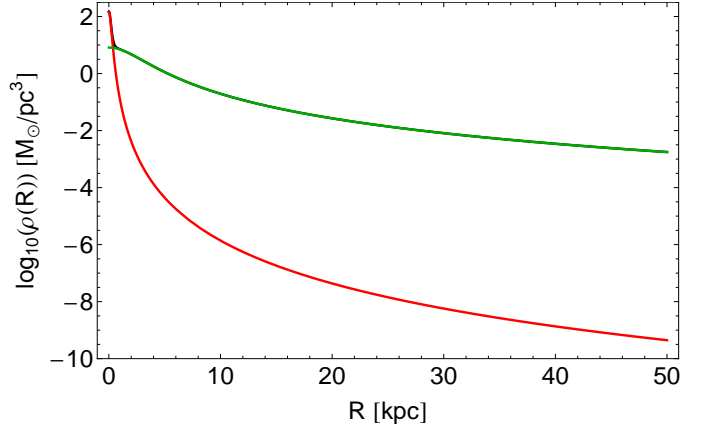}
\caption{Evolution of the mass density $\rho(R)$ in the galactic plane $(z=0)$, as a function of the distance $R$ from the center. We distinguish the contribution of the spherical nucleus (red) and the contribution of the disk (green).}
\label{denevol}
\end{figure}

It is very useful to compute the mass density $\rho(R,z)$ derived from the total potential $\Phi(R,z)$ using the Poisson's equation
\begin{equation}
\rho(R,z) = k \frac{1}{4 \pi G} \nabla^2 \Phi(R,z)
\label{dens}
\end{equation}
In the same equation, we observe the presence of an additional parameter $k = 2.325 \times 10^{-2}$, which is simply a numerical coefficient dictated by the current system of galactic units to obtain the density in units of ${\rm M}_\odot$/pc$^3$. Fig. \ref{denevol}, shows the evolution of the mass density $\rho(R,z=0)$ in the galactic plane as a function of the radius $R$ from the galactic center, where the red line indicates the contribution from the spherical nucleus, while the green line corresponds to the contribution form the disk. It is evident that the density of the nucleus decreases rapidly obtaining very low values, while on the other hand, the density of the disk continues to hold significantly larger values. At large galactocentric distances, the total mass density vary like $1/R^3$ (to be more precise, from the nonlinear fit, we derived that the exact power of the $1/R^n$ asymptotic behavior of the total mass density is $n = 2.977$). This means that the total mass $M(R)$, enclosed in a sphere of radius $R$, increases with the distance. Here, we must point out that our gravitational potential is truncated ar $R_{max} = 50$ kpc, otherwise the total mass of the galaxy modeled by this potential would be infinite, which is obviously not physical.

Exploiting the fact that the $L_z$-component of the total angular momentum is conserved because the gravitational potential $\Phi(R,z)$ is axisymmetric, orbits can be described by means of the effective potential \citep[e.g.,][]{BT08}
\begin{equation}
\Phi_{\rm eff}(R,z) = \Phi(R,z) + \frac{L_z^2}{2R^2}.
\label{veff}
\end{equation}

Then, the basic equations of motion on the meridional plane are
\begin{equation}
\ddot{R} = - \frac{\partial \Phi_{\rm eff}}{\partial R}, \ \ \ \ddot{z} = - \frac{\partial \Phi_{\rm eff}}{\partial z},
\label{eqmot}
\end{equation}
while the equations governing the evolution of a deviation vector ${\bf{w}} = (\delta R, \delta z, \delta \dot{R}, \delta \dot{z})$, which joins the corresponding phase space points of two initially nearby orbits, needed for the
calculation of the standard indicators of chaos (the SALI in our case), are given by the variational equations
\begin{eqnarray}
\dot{(\delta R)} &=& \delta \dot{R}, \ \ \ \dot{(\delta z)} = \delta \dot{z}, \nonumber \\
(\dot{\delta \dot{R}}) &=&
- \frac{\partial^2 \Phi_{\rm eff}}{\partial R^2} \delta R
- \frac{\partial^2 \Phi_{\rm eff}}{\partial R \partial z}\delta z,
\nonumber \\
(\dot{\delta \dot{z}}) &=&
- \frac{\partial^2 \Phi_{\rm eff}}{\partial z \partial R} \delta R
- \frac{\partial^2 \Phi_{\rm eff}}{\partial z^2}\delta z.
\label{vareq}
\end{eqnarray}

Consequently, the corresponding Hamiltonian to the effective potential given in Eq. (\ref{veff}) can be written as
\begin{equation}
H = \frac{1}{2} \left(\dot{R}^2 + \dot{z}^2 \right) + \Phi_{\rm eff}(R,z) = E,
\label{ham}
\end{equation}
where $\dot{R}$ and $\dot{z}$ are momenta per unit mass, conjugate to $R$ and $z$, respectively, while $E$ is the numerical value of the Hamiltonian, which is conserved. Therefore, an orbit is restricted to the area in the meridional plane satisfying $E \geq \Phi_{\rm eff}$.

\section{Orbit classification}
\label{orbclas}

\begin{figure*}[!tH]
\centering
\resizebox{0.9\hsize}{!}{\includegraphics{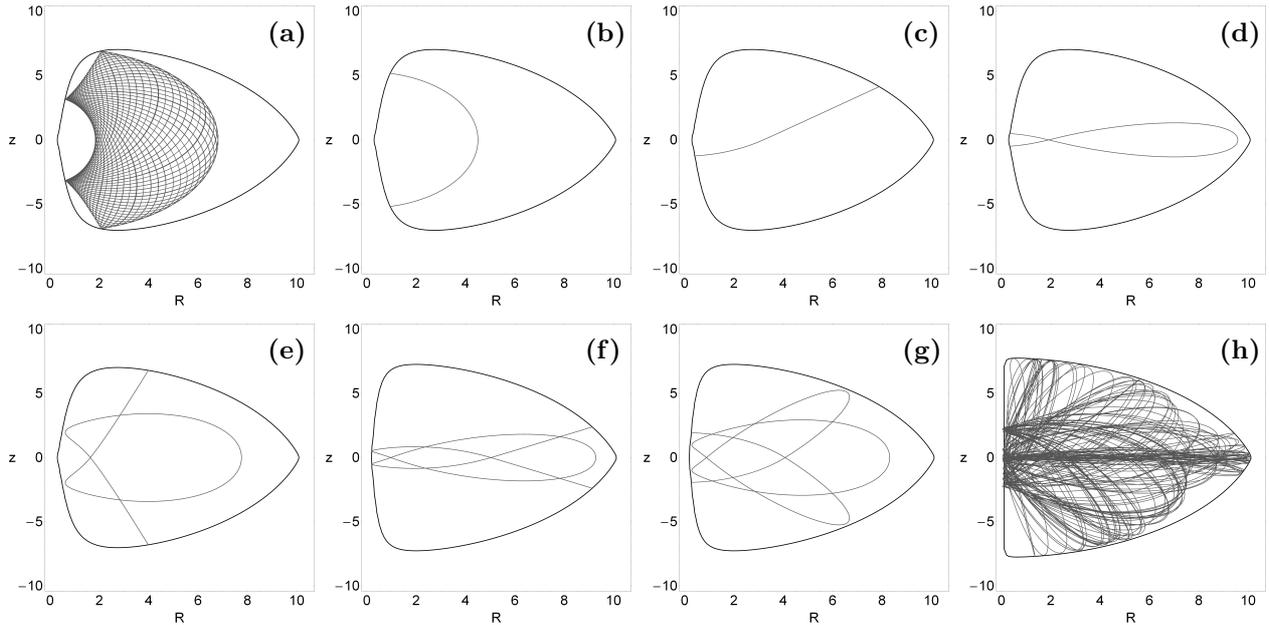}}
\caption{Orbit collection of the basic types of orbits in our galaxy model: (a) box orbit; (b) 2:1 banana-type orbit; (c) 1:1 linear orbit; (d) 2:3 fish-type orbit; (e) 4:3 boxlet orbit; (f) 5:4 boxlet orbit; (g) 6:5 boxlet orbit; (h) chaotic orbit.}
\label{orbs}
\end{figure*}

\begin{table}
\begin{center}
   \caption{Types and initial conditions of the orbits shown in Fig. \ref{orbs}(a-h). In all cases, $z_0 = 0$, $\dot{z_0}$ is found from the energy integral, Eq. (\ref{ham}), while $T_{\rm per}$ is the period of the resonant parent periodic orbits.}
   \label{table1}
   \setlength{\tabcolsep}{3.0pt}
   \begin{tabular}{@{}lcccc}
      \hline
      Figure & Type & $R_0$ & $\dot{R_0}$ & $T_{\rm per}$  \\
      \hline
      \ref{orbs}a &  box         & 1.83000000 &  0.00000000 &           - \\
      \ref{orbs}b &  2:1 banana  & 4.49760475 &  0.00000000 &  1.16485870 \\
      \ref{orbs}c &  1:1 linear  & 3.18361841 & 33.61069890 &  0.92165896 \\
      \ref{orbs}d &  2:3 boxlet  & 9.57565135 &  0.00000000 &  1.87809577 \\
      \ref{orbs}e &  4:3 boxlet  & 7.75468676 &  0.00000000 &  3.55445383 \\
      \ref{orbs}f &  5:4 boxlet  & 8.21911732 &  0.00000000 &  3.65092215 \\
      \ref{orbs}g &  6:5 boxlet  & 8.42887123 &  0.00000000 &  5.23872657 \\
      \ref{orbs}h &  chaotic     & 0.15000000 &  0.00000000 &           - \\
      \hline
   \end{tabular}
\end{center}
\end{table}

In this section, we will numerically integrate several sets of orbits, in an attempt to distinguish the regular or chaotic nature of motion of stars. In all cases, the energy was set equal to $-700$, while the angular momentum of the orbits is treated as a parameter. Here, we have to point out that the energy level controls the size of the grid and particularly $R_{\rm max}$ which is the maximum possible value of the $R$ coordinate. We chose that energy level $(E = -700)$ which yields $R_{\rm max} \simeq 10$ kpc. To study how the angular momentum $L_z$ influences the level of chaos, we let it vary while fixing all the other parameters of our galaxy model. As already said, we fixed the values of all the other parameters and integrate orbits in the meridional plane for the set $L_z = \{1,10,20, ..., 50\}$. Once the values of the parameters were chosen, we computed a set of initial conditions and integrated the corresponding orbits computing the SALI of the orbits and then classifying regular orbits into different families. All the computational methods used for the classification of the orbits are described in detail in \citet{ZC13} and \citet{Z14}.

Our numerical investigation reveals that in our galaxy model there is a plethora of types of orbits: (i) chaotic orbits; (ii) box orbits; (iii) 1:1 linear orbits; (iv) 2:1 banana-type orbits; (v) 2:3 fish-type orbits; (vi) 4:3 resonant orbits; (vii) 5:4 resonant orbit; (viii) 6:5 resonant orbits; and (ix) orbits with other resonances (i.e., all resonant orbits not included in the former categories). It turns out that for these last orbits the corresponding percentage is less than 1\% in all cases, and therefore their contribution to the overall orbital structure of the galaxy is insignificant. A $n:m$ resonant orbit would be represented by $m$ distinct islands of invariant curves in the $(R,\dot{R})$ phase plane and $n$ distinct islands of invariant curves in the $(z,\dot{z})$ surface of section. In Fig. \ref{orbs}(a-h) we present examples of each of the basic types of regular orbits, plus an example of a chaotic one. In all cases, we set $L_z = 20$ (except for the chaotic orbit, where $L_z = 1$). The orbits shown in Figs. \ref{orbs}a and \ref{orbs}h were computed until $t = 100$ time units, while all the parent periodic orbits were computed until one period has completed. The black thick curve circumscribing each orbit is the limiting curve in the meridional plane defined as $\Phi_{\rm eff}(R,z) = E$. Table \ref{table1} shows the types and the initial conditions for each of the depicted orbits; for the resonant cases, the period $T_{\rm per}$ correspond to the parent\footnote{For every orbital family there is a parent (or mother) periodic orbit, that is, an orbit that describes a closed figure. Perturbing the initial conditions which define the exact position of a periodic orbit we generate quasi-periodic orbits that belong to the same orbital family and librate around their closed parent periodic orbit.} periodic orbits.

\begin{figure*}[!tH]
\centering
\resizebox{0.8\hsize}{!}{\includegraphics{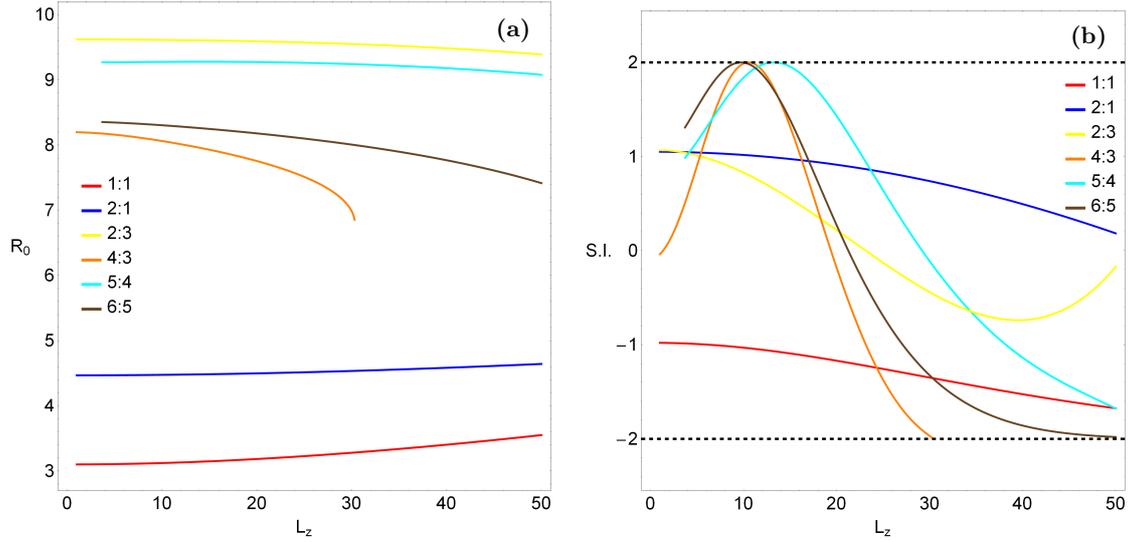}}
\caption{(a-left): The $(R_0,L_z)$ characteristic curves of the orbital families; (b-right): Evolution of the stability index S.I. of the families of periodic orbits shown in Fig. \ref{otd}a. The black horizontal dashed lines at -2 and +2 delimit the range of S.I. for which the periodic orbits are stable.}
\label{otd}
\end{figure*}

In Fig. \ref{otd}a, we present a very informative diagram the so-called ``characteristic" orbital diagram \citep{CM77}. It shows the evolution of the $R$ coordinate of the initial conditions of the parent periodic orbits of each orbital family as a function of the variable angular momentum $L_z$. Here we should emphasize, that for orbits starting perpendicular to the $R$-axis, we need only the initial condition of $R_0$ in order to locate them on the characteristic diagram. On the other hand, for orbits not starting perpendicular to the $R$-axis initial conditions as position-velocity pairs $(R,\dot{R})$ are required and therefore, the characteristic diagram is now three-dimensional providing full information regarding the interrelations of the initial conditions in a tree of families of periodic orbits. Furthermore, the diagram shown in Fig. \ref{otd}b is called the ``stability diagram" \citep{CB85,CM85} and it illustrates the stability of all the families of periodic orbits in our dynamical system when the numerical value of the angular momentum varies, while all the other parameters remain constant. A periodic orbit is stable if only the stability index (S.I.) \citep{Z13} is between −2 and +2. This diagram helps us monitor the evolution of S.I. of the resonant periodic orbits as well as the transitions from stability to instability and vice versa.

\begin{figure*}[!tH]
\centering
\resizebox{0.9\hsize}{!}{\includegraphics{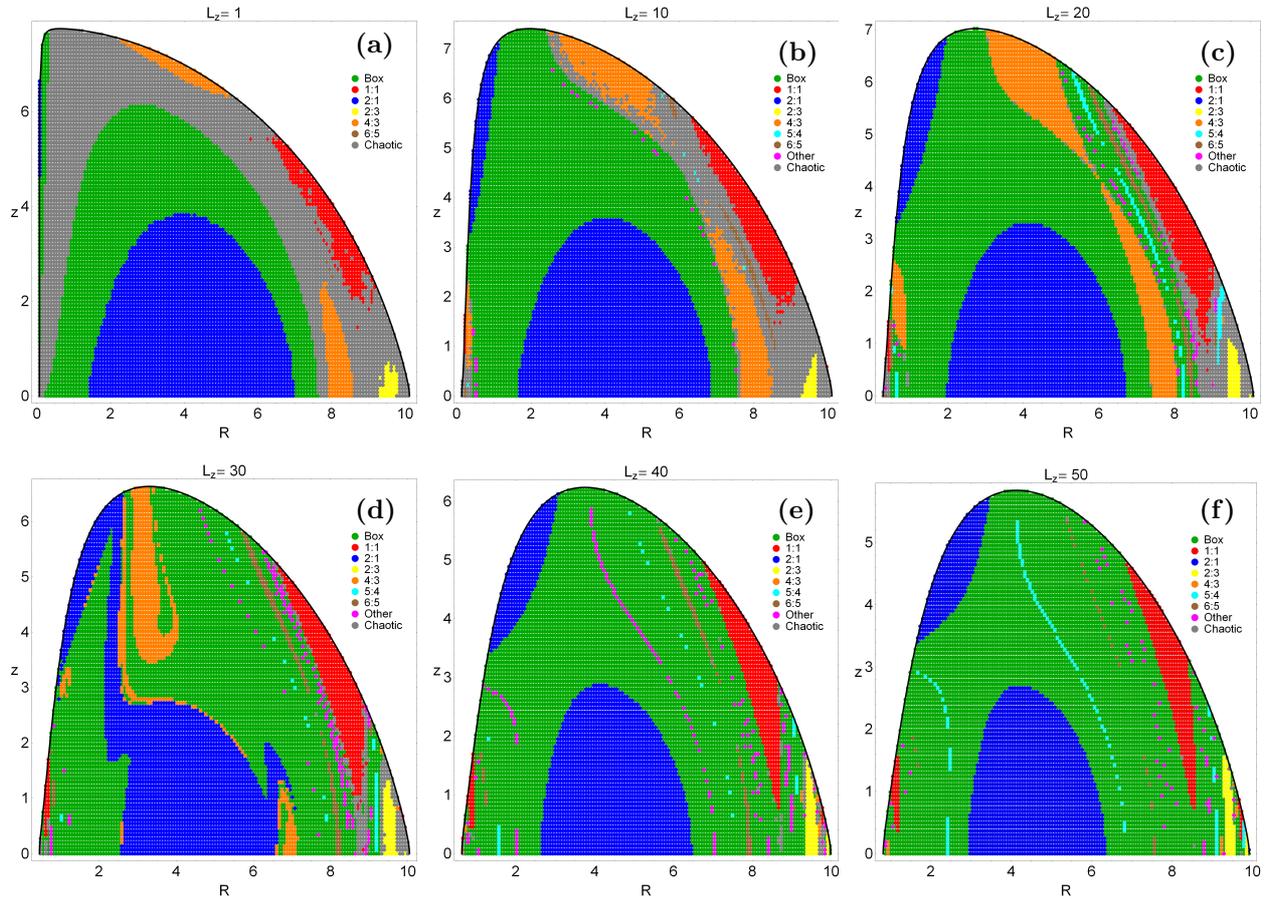}}
\caption{Orbital structure of the physical $(R,z)$ plane of our galaxy model for different values of angular momentum $L_z$.}
\label{grids1}
\end{figure*}

\subsection{The structure of the physical $(R,z)$ space}
\label{pp1}

\begin{figure}[!tH]
\centering
\includegraphics[width=\hsize]{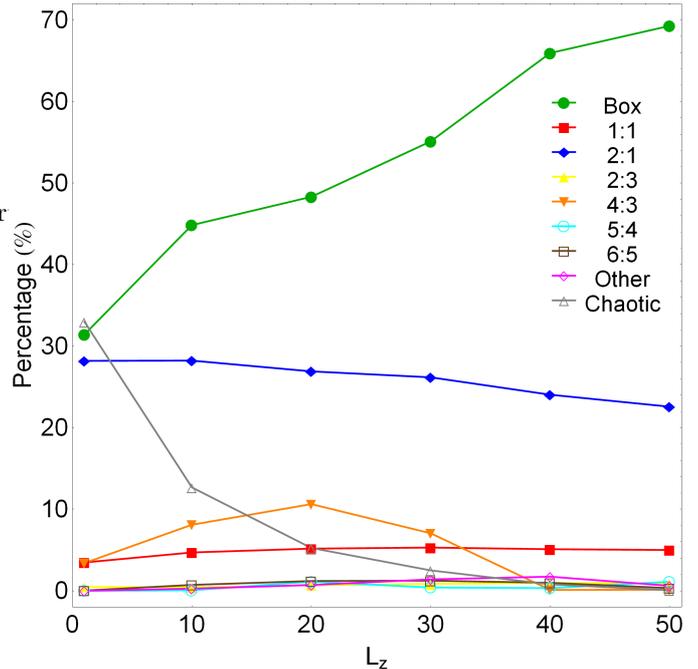}
\caption{Evolution of the percentages of the different types of orbits in the physical $(R,z)$ plane of our galaxy model, when varying the value of the angular momentum $L_z$.}
\label{percs1}
\end{figure}

Our exploration begins in the physical $(R,z)$ plane and in Figs. \ref{grids1}(a-f) we present six grids of initial conditions $(R_0,z_0)$ of orbits that we have classified for different values of the angular momentum $L_z$. For all orbits $\dot{R_0} = 0$, while the initial value of $\dot{z_0}$ is always obtained from the energy integral (\ref{ham}) as $\dot{z_0} = \dot{z}(R_0,\dot{R_0},E) > 0$. The outermost black thick curve circumscribing each grid is the limiting curve in the meridional plane $(R,z)$. Fig. \ref{grids1}a shows the structure of the physical space for $L_z = 1$, that is the case of a model with very low angular momentum stars. We see that the vast majority of the plane is covered by initial conditions corresponding to regular orbits, while initial conditions of chaotic orbits are confined mainly to the outer parts of the grid, where a strong chaotic layer exists. It is evident, that among regular orbits, box orbits is the most populated family, while the 2:1 resonant family is the second most abundant type of regular orbits. We also observe additional smaller stability islands corresponding to 1:1, 2:3 and 4:3 resonant orbits. Our computations suggest that initial conditions of the 6:5 resonant orbits do exit however, the islands of the 6:5 resonant orbits are extremely small, deeply buried in the chaotic sea and therefore, they appear only as lonely points in the grid. It is seen in Fig. \ref{grids1}b that for $L_z = 10$ all the stability islands have been increased in size, thus reducing the area on the physical plane occupied by chaotic orbits. It is interesting to note that an additional 2:1 stability island emerges at the left upper region of the $(R,z)$ plane. Things are quite similar in Fig. \ref{grids1}c where $L_z = 20$ however, the structure of the physical plane starts to change when $L_z = 30$. One may identifies in Fig. \ref{grids1}d two main differences with respect to previous cases: (i) the two 2:1 stability islands have merged and (ii) the 4:3 stability islands appear somehow distorted, indicating that they reach a critical point regarding their stability. Indeed, in Fig. \ref{grids1}e where $L_z = 40$ we see that the 4:3 stability islands disappear, while the 2:1 stability islands separate. Here we have to point out that the absence of the main 4:3 islands for high values of angular momentum was anticipated. In particular, looking the evolution of the characteristic curve of the 4:3 resonance as a function of $L_z$, which is shown in Fig. \ref{otd}, it is seen that this family terminates when $L_z > 30.22$. For higher values of the angular momentum the 4:3 resonance is still present although corresponding only to isolated initial conditions in the physical space but not to well-formed stability islands any more. For high values of angular momentum $(L_z = 50)$ the grid of Fig. \ref{grids1}f shows that almost the entire physical plane is occupied by initial conditions of regular orbits, while chaos, if any, is negligible. Furthermore, we observe that the vast majority of the $(R,z)$ plane is covered by box orbits, while higher resonant orbits (i.e., the 5:4, 6:5, and other resonant families) form thin filaments of initial conditions crossing the extended box area.

The resulting percentages of the chaotic and all types of regular orbits as the value of the angular momentum $L_z$ varies are shown in Fig. \ref{percs1}. It is seen, that box orbits is the most populated family of orbits throughout the values of $L_z$. In galaxy models with very low angular momentum stars $(L_z = 1)$ we observe that the percentages of box and chaotic orbits coincide at about 33\%, thus sharing about two thirds of the physical plane. As we proceed to higher angular momentum models the percentage of chaotic orbits reduces rapidly and for $L_z > 50$ it vanishes. In contrast, the percentage of box orbits increases almost linearly and at the highest studied value of the angular momentum box orbits cover about 70\% of the physical $(R,z)$ plane. On the other hand, the 1:1 and 2:1 resonant families seem to be little affected by the change of the angular momentum. In particular, the percentage of 2:1 orbits exhibits a small reduction with increasing $L_z$, while at the same time, the rate of the 1:1 resonant orbits seems to increases slightly. It is seen, that the evolution of the percentage of the 4:3 resonant family is very interesting. Indeed, this resonance has a strong presence only for $1 \leq L_z \leq 20$ occupying up to about one tenth of the physical plane, while for larger value of the angular momentum $(L_z > 30)$ it becomes unstable thus having a very low percentage. Furthermore, it is evident that in general, all the remaining types of resonant orbits (i.e., the 2:3, 5:4, 6:5, and other higher resonances) remain almost unperturbed possessing very low percentages throughout (less than 5\%). Therefore, one may reasonably conclude that in the physical $(R,z)$ plane box, 4:3, and chaotic orbits are those mostly affected by the change of the value of the angular momentum.

\begin{figure*}[!tH]
\centering
\resizebox{0.9\hsize}{!}{\includegraphics{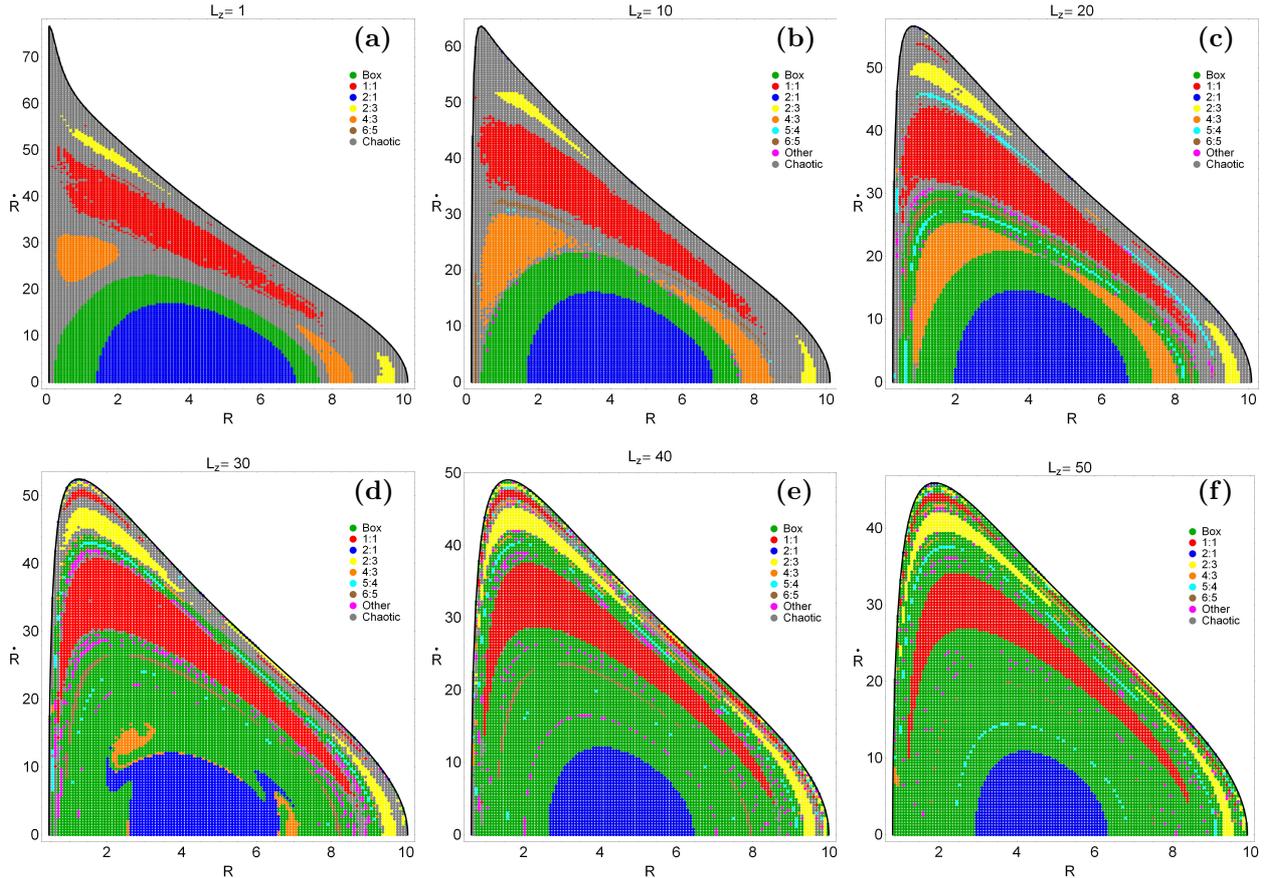}}
\caption{Orbital structure of the phase $(R,\dot{R})$ plane of our galaxy model for different values of angular momentum $L_z$.}
\label{grids2}
\end{figure*}

\subsection{The structure of the phase $(R,\dot{R})$ space}
\label{pp2}

\begin{figure}[!tH]
\centering
\includegraphics[width=\hsize]{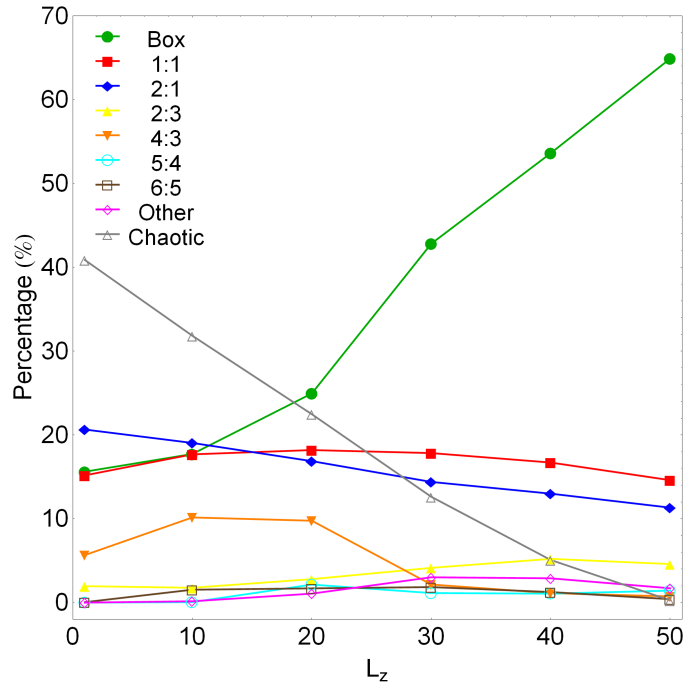}
\caption{Evolution of the percentages of the different types of orbits in the phase $(R,\dot{R})$ plane of our galaxy model, when varying the value of the angular momentum $L_z$.}
\label{percs2}
\end{figure}

We continue our investigation in the phase $(R,\dot{R})$ space and in Figs. \ref{grids2}(a-f) we present six grids of initial conditions $(R_0,\dot{R_0})$ of orbits that we have classified for different values of the angular momentum $L_z$. All orbits have $z_0 = 0$, while again the initial value of $\dot{z_0}$ is obtained from the energy integral (\ref{ham}). The outermost black thick curve is the limiting curve which is defined as
\begin{equation}
\frac{1}{2} \dot{R}^2 + \Phi_{\rm eff}(R,0) = E.
\label{zvc}
\end{equation}
We see in Fig. \ref{grids2}a, that when the angular momentum is very low $(L_z = 1)$, an extended chaotic sea covers the majority of the phase plane, while several stability islands corresponding to different types of resonant orbits are present and mainly embedded in the chaotic domain. However, as the value of the angular momentum increases the structure of the phase plane changes significantly\footnote{It should be pointed out that the permissible area on the physical as well as on the phase plane is reduced as we increase the value of the angular momentum.}. In Figs. \ref{grids2}(b-c) where the value of the angular momentum is 10 and 20, respectively, the only observable difference is the decrease of the chaotic area thus giving place to all stability islands to increase their size. The drastic changes start to appear for $L_z \geq 30$, where in Fig. \ref{grids2}d we see, first of all, the deformation of the 4:3 stability islands which has been explained previously when studying the physical plane (for $L_z > 30$ initial conditions corresponding to 4:3 resonant orbits are still present located mainly at the outer parts of the plane, although the main stability islands disappear). In galactic models where stars possess high enough angular momentum (see Figs. \ref{grids2}(e-f) where the value of $L_z$ is 40 and 50, respectively) regular orbits occupy almost all the available phase plane, while chaotic motion is negligible. Moreover, the grids are very rich in types of orbits, taking into account that several families of higher resonant orbits are present producing tiny chains of stability islands mainly inside the box area. It should also be mentioned that box orbits grow in expense of other important orbital families as the extent of the stability islands of the two basic types of orbits, that is the 1:1 and the 2:1 resonant families, seems to be reduced with increasing angular momentum.

\begin{figure*}[!tH]
\centering
\resizebox{0.9\hsize}{!}{\includegraphics{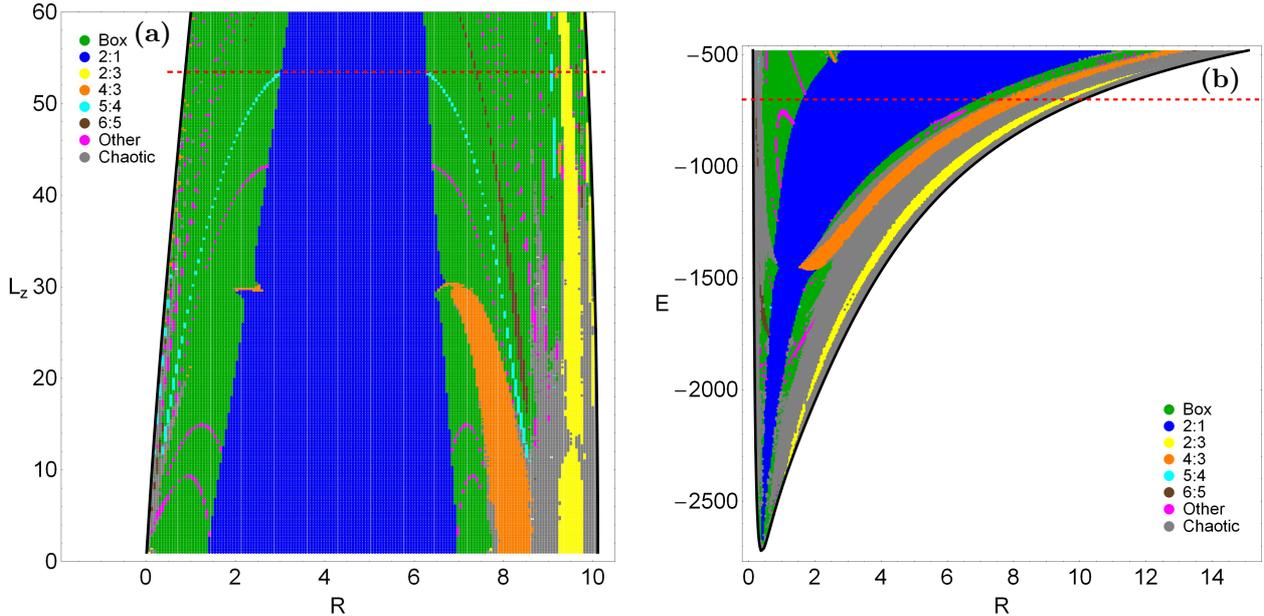}}
\caption{Orbital structure of the (a-left): $(R,L_z)$-plane and (b-right): $(R,E)$-plane when $L_z = 10$. These diagrams give a detailed analysis of the evolution of orbits starting perpendicularly to the $R$-axis when the value of (a) the angular momentum and (b) the total orbital energy varies.}
\label{grd}
\end{figure*}

The following Fig. \ref{percs2} shows the evolution of the percentages of the chaotic and all types of regular orbits as a function of the angular momentum $L_z$. One may observes that as we proceed to higher values of angular momentum the percentage of chaotic orbits decreases almost linearly, while at the same time, the rate of box orbits grows steadily. In the case of very low angular momentum $(L_z = 1)$ about 40\% of the phase plane corresponds to initial conditions of chaotic orbits and only 15\% of it to box orbits however, when $L_z > 20$ box orbits is the most populated family dominating the phase plane. At the highest studied value of the angular momentum $(L_z = 50)$, the entire phase plane is covered only by initial conditions of regular orbits (about 70\% are box orbits), while chaos is practically absent (our computations indicate an extremely low rate of about 0.3\%). Furthermore, our numerical analysis suggests that all the remaining families of orbits are significantly less affected by the change in the value of the angular momentum. In particular, the percentages of the 2:1 meridional banana-type orbits and also that of the 1:1 linear orbits are little influenced by the increase of the angular momentum. It is also seen, that when $L_z > 20$ the percentage of the 4:3 resonant family drops suddenly, remaining at very low values from then on. In addition, we may say that in general terms, all the other resonant families (i.e., the 2:3, 5:4, 6:5, and other families) hold throughout very low percentages (always less than 5\%), so varying the value of $L_z$ only shuffles the orbital content among them. Thus, taking into account all the above-mentioned analysis we may argue that in the phase $(R,\dot{R})$ plane the types of orbits that are mostly influenced by the angular momentum are the box, 4:3, and chaotic orbits.

\subsection{An overview analysis}
\label{geno}

The grids in physical $(R,z)$ as well as in the phase $(R,\dot{R})$ plane can provide information on the phase space mixing for only a fixed value of the angular momentum $L_z$. H\'{e}non however, back in the 60s \citep{H69}, considered a plane which provides information about regions of regularity and regions of chaos using the section $z = \dot{R} = 0$, $\dot{z} > 0$, i.e., the test particles (stars) launched on the $R$-axis, parallel to the $z$-axis and in the positive $z$-direction. Thus, in contrast to the previously discussed grids, only orbits with pericenters on the $R$-axis are included and therefore, the value of $L_z$ is used as an ordinate. Fig. \ref{grd}a shows the orbital structure of the $(R,L_z)$-plane when $L_z \in [1, 60]$. In order to be able to monitor with sufficient accuracy and details the evolution of the families of orbits, we defined a dense grid of $10^5$ initial conditions in the $(R,L_z)$-plane. It is evident, that the vast majority of the that grid is covered either by box or 2:1 resonant orbits, while initial conditions of chaotic orbits are mainly confined to right outer part of the $(R,L_z)$-plane. This diagram shows clearly that the main stability islands of the 4:3 resonance cease to exits when $L_z > 30.2176$ which is very close to the termination value obtained from the characteristic curve of the diagram shown in Fig. \ref{otd}. It is also seen, that several families of higher resonant orbits are present corresponding to thin filaments of initial conditions living inside the box region. Our numerical calculations revealed that apart from the 5:4 and 6:5 families, a large collection of higher secondary resonant orbits has been identified (i.e., 4:7, 7:5, 7:6, 8:7, 8:9, 9:7, 10:7 and 10:9).

It should be emphasized, that the minimum value of the $R$ coordinate $(R_{min})$ is reduced with increasing $L_z$. This is an issue of great importance because in \citet{ZC13,Z14} we proved that a necessary condition for an orbit to be chaotic is to pass near the center of the potential. Therefore, only low angular momentum stars can approach very close to the galactic center (or in other words, near the massive nucleus) thus receiving a large acceleration and exhibit chaotic motion. On the other hand, high angular momentum stars are not allowed to come close to the central nucleus and this explains why the majority of high angular momentum stars move in regular orbits, while chaotic motion is very limited. The horizontal red dashed line at $L_z = 53.44$ marks the last indication of chaos; for $L_z > 53.44$ the motion of stars is entirely regular and there is no evidence of chaotic motion whatsoever. In fact, $L_z = 53.44$ is the maximum value of the angular momentum for which the orbits can display chaotic motion and it is called the critical value of the angular momentum $L_{z\rm c}$ \citep[e.g.,][]{Z11,Z12,ZC13}. Here we must stress out that the $(R,L_z)$-plane contains only such orbits starting perpendicularly to the $R$-axis, while orbits whose initial conditions are pairs of position-velocity (i.e., the 1:1 resonant family) are obviously not included.

Our previous experience \citep[e.g.,][]{ZC13,Z14} indicates that one of the most important parameters that influences greatly the orbital structure of galaxies is the isolated integral of the total orbital energy $E$. In our research we had to fix the energy level at -700 in order to maintain a constant $R_{max}$ at about 10 kpc, when varying the value of the angular momentum. Nevertheless, we decided to perform some additional calculations in order to explore how the total orbital energy affects the nature of stars. Our results are shown in Fig. \ref{grd}b where we followed a similar approach to that explained previously in Fig. \ref{grd}a varying now the value of the orbital energy in the interval $E \in [-2720, -480]$ when $L_z = 10$. We chose that particular value of the angular momentum $(L_z = 10)$ so that there is a sufficient amount of chaos in the system. We observe that for very low values of the energy $(E < -2380)$, that is the case of local motion in galaxy models, stars move either in box, 2:1 or chaotic orbits. On the other hand, higher resonant orbits appear only at larger energy levels (i.e., the 4:3 resonance emerges for $E > 1460$). As in Fig. \ref{grd}a, a plethora of higher secondary resonant orbits (i.e., 2:5, 5:8, 7:5, 7:6, 8:7, 9:7, 9:8, 9:10, 10:7) has been identified when varying the orbital energy. It is also clear, that the portion of the 2:1 resonant orbits grows rapidly with increasing energy and when $E > 1000$, the 2:1 meridional banana-type orbits is the most populated family. The horizontal red dashed line denotes the constant energy level $E = -700$ in which all the previous computations took place.

\section{Conclusions and discussion}
\label{disc}

The aim of the present work was to investigate how influential is the angular momentum on the level of chaos and on the distribution of regular families among its orbits. For this purpose, we used an analytic, axially symmetric galactic gravitational model which embraces the general features of a disk galaxy with a dense, massive, central nucleus. To simplify our study, we chose to work in the meridional plane $(R,z)$, thus reducing three-dimensional to two-dimensional motion. We kept the values of all the other parameters constant, because our main objective was to determine the influence of the angular momentum on the percentages of the orbits. Our thorough and detailed numerical analysis suggests that the level of chaos as well as the distribution in regular families is indeed very dependent on the angular momentum of stars. Here, we should point out that the present article is the last part of a series of papers \citep{ZC13,CZ13,ZCar13,ZCar14} that have as their main objective the orbit classification (not only regular versus chaotic, but also separating regular orbits into different regular families) in different galactic gravitational potentials. Thus, we decided to follow a similar structure and of course the same numerical approach in all of them.

In earlier papers of the series \citep[e.g.,][]{ZC13,Z14}, we demonstrated how the dynamical parameters of the system, such as the mass of the nucleus, the mass of the disk, the scale length of nucleus, etc influence the orbital structure of galaxies. The present investigation takes place in the physical $(R,z)$ and also in the phase $(R,\dot{R})$ space for a better understanding of the orbital structure of the system. To show how the angular momentum influences the orbital structure of the system, we presented for each case, dense color-coded grids of about 50000 initial conditions, which allow us to visualize what types of orbits occupy specific areas in the physical/phase space. Each orbit was numerically integrated for a time interval of $10^4$ time units ($10^{12}$ yr), which corresponds to a time span of the order of hundreds of orbital periods. The particular choice of the total integration time is an element of great importance, especially in the case of the sticky orbits. The main numerical outcomes of our research can be summarized as follows:
\begin{enumerate}
 \item Several types of regular orbits were found to exist in our galactic gravitational model, while there are also extended chaotic domains separating the areas of regularity. In particular, a large variety of resonant orbits (i.e., 1:1, 2:1, 2:3, 4:3, 5:4, 6:5, and higher resonant orbits) are present, thus making the orbital structure more rich.
 \item It was found that in both the physical $(R,z)$ and the phase $(R,\dot{R})$ space the angular momentum influences mainly box, 4:3, and chaotic orbits. Moreover, the majority of starts move in regular orbits and in general terms, box orbits is the most populated family throughout the range of values of $L_z$.
 \item The largest amount of chaos was measured for low values of $L_z$ $(L_z < 10)$ corresponding to low angular momentum stars, while as $L_z$ increases, the amount of chaotic orbits is reduced rapidly and for high enough values of the angular momentum $(L_z > 40)$, almost all tested orbits were found to be regular.
 \item The drastic decrease of the observed chaos with increasing $L_z$ was justified taking into account the fact the high angular momentum stars are not allowed to approach very close to the central massive nucleus thus encountering strong forces and exhibiting chaotic motion as low angular momentum stars do.
 \item One of the most influential parameters of the dynamical system is undoubtedly the total orbital energy. Indeed, it was observed that for low energy levels corresponding to local motion near the nucleus, the orbital content is rather poor since most of the resonant orbits emerge at relatively high values of energy suitable for global motion of stars.
\end{enumerate}

We consider the results of the present research as an initial effort and also as a promising step in the task of exploring the orbital structure of disk galaxies with a central and spherical nucleus. Taking our encouraging outcomes into account, it is in our future plans to properly modify our galaxy model in order to expand our investigation into three dimensions. This will allow us to unveil how the basic parameters entering the system influence the nature of three-dimensional orbits. Also, we would be particularly interested in revealing the evolution of the percentages of all the families of orbits when varying the different dynamical quantities of the galactic model.

\section*{Acknowledgments}

I would like to express my warmest thanks to Dr. D.D. Carpintero for all the illuminating and creative discussions during this research and of course for his substantial contribution to our efforts to refine and improve even further the orbit classification code. My thanks also go to the anonymous referee for the careful reading of the manuscript and for all the apt comments which allowed us to improve the clarity of the paper.


\section*{References}

\begin{thebibliography}{}

\bibitem[\protect\citeauthoryear{Allen \& Santill\'an}{1991}]{AS91} Allen, C., Santill\'an, A., 1991. An improved model of the galactic mass distribution for orbit computations. Rev. Mex. 22, 255-263.

\bibitem[\protect\citeauthoryear{Binney \& Tremaine}{2008}]{BT08} Binney, J., Tremaine, S., 2008. Galactic Dynamics, Princeton Univ. Press, Princeton, USA.

\bibitem[\protect\citeauthoryear{Bountis et al.}{2012}]{BMA12} Bountis, T., Manos, T., Antonopoulos, Ch., 2008. Complex statistics in Hamiltonian barred galaxy models. Celest. Mech. Dyn. Astron. 113, 63-80.

\bibitem[\protect\citeauthoryear{Caranicolas \& Vozikis}{1986}]{CV86} Caranicolas, N.D, Vozikis, Ch., 1986. Orbital characteristics of dynamical models of elliptical galaxies. Cel. Mech. 39, 85-102.

\bibitem[\protect\citeauthoryear{Caranicolas \& Zotos}{2013}]{CZ13} Caranicolas, N.D., Zotos, E.E., 2013. Unveiling the influence of dark matter in axially symmetric galaxies. Pub. Ast. Soc. Australia 30, 49-63.

\bibitem[\protect\citeauthoryear{Carpintero \& Aguilar}{1998}]{CA98} Carpintero, D.D., Aguilar, L.A., 1998. Orbit classification in arbitrary 2D and 3D potentials. Mon. Not. R. Astron. Soc. 298, 1-21.

\bibitem[\protect\citeauthoryear{Contopoulos}{1960}]{C60} Contopoulos, G., 1960. A third integral of motion in a galaxy. Z. Astroph. 49, p273.

\bibitem[\protect\citeauthoryear{Contopoulos}{1979}]{C79} Contopoulos, G., 1979. In Stochastic behavior in classical and quantum Hamiltonian systems, G. Casati and J. Ford Eds., p. 1--17.

\bibitem[\protect\citeauthoryear{Contopoulos \& Barbanis}{1985}]{CB85} Contopoulos, G., Barbanis, B., 1985. 	
	Resonant systems with three degrees of freedom. Astron. Astrophys. 153, 44-54.

\bibitem[\protect\citeauthoryear{Contopoulos \& Grosb{\o}l}{1989}]{CG89} Contopoulos, G., Grosb{\o}ol, P., 1989.
	Orbits in barred galaxies. Astron. Astrophys. Rev. 1, 261-289.

\bibitem[\protect\citeauthoryear{Contopoulos \& Magnenat}{1985}]{CM85} Contopoulos, G., Magnenat, P., 1985. Simple three-dimensional periodic orbits in a galactic-type potential. Cel. Mech. 37, 387-414.

\bibitem[\protect\citeauthoryear{Contopoulos \& Mertzanides}{1977}]{CM77} Contopoulos, G., Mertzanides, C, 1977. Inner Lindblad resonance in galaxies: Nonlinear theory. II - Bars. Astron. Astrophys. 61, 477-485.

\bibitem[\protect\citeauthoryear{Copin et al.}{2000}]{CZD00} Copin, Y., Zhao, H., de Zeeuw, P., 2000.
	Probing a regular orbit with spectral dynamics. Mon. Not. R. Astron. Soc. 318, 781-797.

\bibitem[\protect\citeauthoryear{Gerhard \& Saha}{1991}]{GS91} Gerhard, O., Saha, P., 1991. 	
	Recovering galactic orbits by perturbation theory. Mon. Not. R. Astron. Soc. 251, 449-467.

\bibitem[\protect\citeauthoryear{Greiner}{1987}]{G87} Greiner, J., 1987. A new kind of stellar orbit in a galactic potential. Cel. Mech. 40, 171-175.

\bibitem[\protect\citeauthoryear{H\'{e}non}{1969}]{H69} H\'{e}non, M., 1969. Numerical exploration of the restricted problem. Astron. Astrophys. 1, 223-238.

\bibitem[\protect\citeauthoryear{Karanis \& Caranicolas}{2001}]{KC01} Karanis, G.I., Caranicolas, N.D., 2001. Transition from regular motion to chaos in a logarithmic potential. Astron. Astrophys. 367, 443-448.

\bibitem[\protect\citeauthoryear{Kaufmann \& Patsis}{2005}]{KP05} Kaufmann, D., Patsis, P., 2005. Propeller orbits in barred galaxy models. Astrophys. J. 624, 693-700.

\bibitem[\protect\citeauthoryear{Lees \& Schwarzschild}{1992}]{LS92} Lees, J.F., Schwarzschild, M., 1992. The orbital structure of galactic halos. Astrophys. J. 384, 491-501.

\bibitem[\protect\citeauthoryear{Manabe}{1979}]{M79} Manabe, S., 1979. Applicability of approximate third integral of motion for stellar orbits in the galaxy. Pub. Ast. Soc. Japan 31, 369-394.

\bibitem[\protect\citeauthoryear{Manos \& Athanassoula}{2011}]{MA11} Manos, T., Athansssoula, E., 2011. Regular and chaotic orbits in barred galaxies - I. Applying the SALI/GALI method to explore their distribution in several models. Mon. Not. R. Astron. Soc. 415, 629-642.

\bibitem[\protect\citeauthoryear{Manos et al.}{2013}]{MBS13} Manos, T., Bountis, T., Skokos, Ch., 2013. Interplay between chaotic and regular motion in a time-dependent barred galaxy model. J. Phys. A: Math. Theor. 46, 254017.

\bibitem[\protect\citeauthoryear{Martinet \& Mayer}{1975}]{MM75} Martinet, L, Mayer, F., 1975. Galactic orbits and integrals of motion for stars of old galactic populations. III - Conclusions and applications. Astron. Astrophys. 44, 45-57.

\bibitem[\protect\citeauthoryear{Miyamoto \& Nagai}{1975}]{MN75} Miyamoto, W., Nagai, R., 1975. Three-dimensional models for the distribution of mass in galaxies. Pub. Ast. Soc. Japan 27, 533-543.

\bibitem[\protect\citeauthoryear{Oll\'{e} \& Pfenniger}{1998}]{OP98} Oll\'{e}, M., Pfenniger, D., 1998. Vertical orbital structure around the Lagrangian points in barred galaxies. Link with the secular evolution of galaxies. Astron. Astrophys. 334, 829-839.

\bibitem[\protect\citeauthoryear{Ollongren}{1965}]{O65} Ollongren, A., 1965. Theory of stellar orbits in the galaxy. Ann. Rev. Astron. Astrophys. 3, 113-134.

\bibitem[\protect\citeauthoryear{Ollongren}{1967}]{O67} Ollongren, A., 1967. Construction of galactic stellar orbits similar to harmonic oscillators. Astron. J. 72, 436-442.

\bibitem[\protect\citeauthoryear{Pfenniger}{1984}]{P84} Pfenniger, D., 1984. The 3D dynamics of barred galaxies. Astron. Astrophys. 134, 373-386.

\bibitem[\protect\citeauthoryear{Pfenniger}{1996}]{P96} Pfenniger, D., 1996. In Buta R., Crocker D. A., Elmegreen B. G., eds, ASP Conf. Ser. Vol. 91, Barred Galaxies. Astron. Soc. Pac., San Francisco, p. 273.

\bibitem[\protect\citeauthoryear{Pichardo et al.}{2004}]{PMM04} Pichardo, B., Martos, M., Moreno, E., 2004. Models for the gravitational field of the galactic bar: An application to stellar orbits in the galactic plane and orbits of some globular clusters. Astrophys. J. 609, 144-165.

\bibitem[\protect\citeauthoryear{Sellwood \& Wilkinson}{1993}]{SW93} Sellwood, J., Wilkinson, A., 1993. Dynamics of barred galaxies. Rep. Prog. Phys. 56, 173-256.

\bibitem[\protect\citeauthoryear{Skokos et al.}{2002a}]{SPA02a} Skokos, Ch., Patsis, P.A., Athanassoula, E., 2002a. Orbital dynamics of three-dimensional bars - I. The backbone of three-dimensional bars. A fiducial case. Mon. Not. R. Astron. Soc. 333, 847-860.

\bibitem[\protect\citeauthoryear{Skokos et al.}{2002b}]{SPA02b} Skokos, Ch., Patsis, P.A., Athanassoula, E., 2002b. Orbital dynamics of three-dimensional bars - II. Investigation of the parameter space. Mon. Not. R. Astron. Soc. 333, 861-870.

\bibitem[\protect\citeauthoryear{Zotos}{2011}]{Z11} Zotos, E.E., 2011. A new dynamical model for the study of galactic structure. New Astron. 16, 391-404.

\bibitem[\protect\citeauthoryear{Zotos}{2012}]{Z12} Zotos, E.E., 2012. Exploring the nature of orbits in a galactic model with a massive nucleus. New Astron. 17, 576-588.

\bibitem[\protect\citeauthoryear{Zotos}{2013}]{Z13} Zotos, E.E., 2013. Revealing the evolution, the stability and the escapes of families of resonant periodic orbits in Hamiltonian systems. Nonlin. Dyn. 73, 931-962.

\bibitem[\protect\citeauthoryear{Zotos}{2014}]{Z14} Zotos, E.E., 2014. Classifying orbits in galaxy models with a prolate or an oblate dark matter halo component. Astron. Astrophys. 563, A19.

\bibitem[\protect\citeauthoryear{Zotos \& Carpintero}{2013}]{ZC13} Zotos, E.E., Carpintero, D.D., 2013. Orbit classification in the meridional plane of a disk galaxy model with a spherical nucleus. Celest. Mech. Dyn. Astron. 116, 417-438 (Paper I).

\bibitem[\protect\citeauthoryear{Zotos \& Caranicolas}{2013}]{ZCar13} Zotos, E.E., Caranicolas, N.D., 2013. Revealing the influence of dark matter on the nature of motion and the families of orbits in axisymmetric galaxy models. Astron. Astrophys. 560, A110.

\bibitem[\protect\citeauthoryear{Zotos \& Caranicolas}{2014}]{ZCar14} Zotos, E.E., Caranicolas, N.D., 2014. Determining the nature of orbits in disk galaxies with non spherical nuclei. Nonlin. Dyn. 76, 323-344.

\end{thebibliography}
\end{document}